\newif\iffigs\figstrue
\documentclass[paper, 12pt, letterpaper, epsf]{JHEP}
\input{epsf.tex}
\usepackage{graphics}
\def\Bbb{\bf}
\def\C{{\Bbb C}}
\def\R{{\Bbb R}}
\def\Z{{\Bbb Z}}

\def\bearray{\begin{eqnarray}}
\def\eearray{\end{eqnarray}}
\def\bearraynn{\begin{eqnarray*}}
\def\eearraynn{\end{eqnarray*}}
\def\bfig{\begin{figure}}
\def\efig{\end{figure}}

\def\opeq#1{\advance\lineskip#1 \advance\baselineskip#1
        \advance\lineskiplimit#1}

\newtheorem{Proposition}{Proposition}[section]

\newtheorem{Theorem}{Theorem}[section]
\newtheorem{Lemma}{Lemma}[section]
\newtheorem{Corrolary}{Corrolary}[section]

\newcommand{\be}{\begin{equation}}
\newcommand{\ee}{\end{equation}}
\newcommand{\bea}{\begin{eqnarray}}
\newcommand{\eea}{\end{eqnarray}}

\newcommand{\bp}{\begin{Proposition}}
\newcommand{\ep}{\end{Proposition}}
\newcommand{\bt}{\begin{Theorem}}
\newcommand{\et}{\end{Theorem}}
\newcommand{\bl}{\begin{Lemma}}
\newcommand{\el}{\end{Lemma}}
\newcommand{\bc}{\begin{Corrolary}}
\newcommand{\ec}{\end{Corrolary}}
\newcommand{\nn}{\nonumber}

\font\mybb=msbm10 at 12pt

\def\bb#1{\hbox{\mybb#1}}

\def\Z {\bb{Z}}

\def\id{\protect{{1 \kern-.28em {\rm l}}}}

\def\bea{\begin{eqnarray}}
\def\eea{\end{eqnarray}}

\def\k{{\bf k}}

\def\be{\begin{equation}}
\def\ee{\end{equation}}
\def\ba{\begin{eqnarray}}
\def\ea{\end{eqnarray}}

\def\1{{{(1)}}}\def\2{{{(2)}}}\def\3{{{(3)}}}

\font\mybb=msbm10 at 12pt

\def\bb#1{\hbox{\mybb#1}}

\def\Z {\bb{Z}}

\def\id{\protect{{1 \kern-.28em {\rm l}}}}

\usepackage{graphics}

\title{An analytic torsion for graded D-branes}

\author{C.~I.~Lazaroiu
\\C.~N.~Yang Institute for Theoretical Physics\\
SUNY at Stony BrookNY11794-3840,
U.S.A.\\calin@insti.physics.sunysb.edu}

\abstract{I consider the semiclassical approximation of the 
graded Chern-Simons field theories describing certain systems of topological A
type branes in the large radius limit of Calabi-Yau compactifications.
I show that the semiclassical 
partition function can be expressed in terms of a 
certain (differential) numerical invariant which is 
a version of the analytic torsion of Ray and Singer, 
but associated with flat graded superbundles. I also discuss a `twisted' version 
of the Ray-Singer norm, and show its independence of metric data.
As illustration, I consider
graded D-brane pairs of unit relative grade with a scalar condensate in the 
boundary condition changing sector. 
For the particularly simple case when the reference flat connections
are trivial, I show that the generalized torsion reduces to a 
power of the classical Ray-Singer invariant of the base 3-manifold. }

\preprint{YITP-SB 01-62}

\begin{document}

\tableofcontents

\pagebreak
\vskip .6in

\section{Introduction}

D-brane composite formation is a subject of central importance for 
a deeper understanding of open string theory dynamics. Of particular 
interest is the incarnation of such processes for the case of superstring 
compactifications on Calabi-Yau backgrounds, which provide a rich source 
of potential phenomenological applications. 

A systematic study of D-brane composites 
is best performed with the tools of string field theory. Indeed, such 
processes involve {\em off-shell} string dynamics, which is best captured
in string field language. 
Given the complexity superstring models, 
a satisfactory formulation of superstring field theory is 
lacking Calabi-Yau backgrounds. However, part of the dynamics of 
spacetime fields originating from the chiral primary sector allows for an 
explicit  description in terms of topological strings\cite{Witten_nlsm,
Witten_mirror, Witten_CS}. 

A fundamental observation made in \cite{Douglas_Kontsevich}(see also 
\cite{Zaslow_Polishchuk} and \cite{Seidel}) is that 
D-branes in Calabi-Yau backgrounds are graded objects. In our context,
this means that topological D-branes will 
carry certain integral data which specify a branch of their 
`BPS grade', which was discussed at length in \cite{Douglas_Kontsevich}.
It was recently proposed \cite{com1, com3, Diaconescu, sc, bv} that the 
dynamics of graded topological D-branes is described
by certain target space theories which have the form of  
`graded Chern-Simons models'\footnote{For the A-model, these 
theories do not take into account worldsheet instanton corrections
(which can be formally 
incorporated along the lines of \cite{Fukaya, Fukaya2}).
In the present paper, we always work in the large radius limit
of a given Calabi-Yau compactification, where such corrections 
can be neglected.}. These are versions of Chern-Simons field 
theories associated with graded superbundles, and whose fields assemble into a 
`graded superconnection of total degree one'.
This approach, which is intimately connected with the derived category 
program of \cite{Kontsevich} and \cite{Douglas_Kontsevich, Aspinwall, 
Douglas_Aspinwall}, allows  for 
a description of topological D-brane physics in standard field theoretic terms.
It also leads to a physical representation of the extended moduli space of 
open strings \cite{bv, gauge}. 

One benefit of this description is that the formal analogy between graded 
and ungraded Chern-Simons theories suggests a wealth of generalizations 
of classical connections between physics and certain mathematical invariants. 
In this note, we take a first step in this direction for the graded 
Chern-Simons theory of A-type branes.

It is well-known that the semiclassical approximation of 
usual Chern-Simons theory (which forms the string 
field theory of single, ungraded topological A-branes) is 
related to the analytic torsion of Ray and Singer. 
In fact, the Chern -Simons approach 
allows one to `discover' the torsion from physical considerations. 
It is natural to ask what is the analogue of this relation for 
graded Chern-Simons models. Proceeding in physical manner, we  
consider the semiclassical approximation of graded Chern-Simons theories, 
from which we extract a generalized version of Ray-Singer torsion, 
which should prove useful for a topological characterization of 
D-brane composites in Calabi-Yau backgrounds.

The note is organized as follows. In Section 2, we give a short review 
of the graded Chern-Simons theory of \cite{sc, bv} and 
recall the D-brane interpretation of its vacuum configurations. 
In Section 3, we consider the semiclassical approximation 
to these models around a general background. 
Upon using the methods of \cite{ST, Schwarz_resolvent, Adams_Sen} and 
related work, we evaluate the partition function in this limit 
up to a real prefactor related to the isotropy subgroup 
of the background. This allows us to express the result in terms of 
a certain generalization of the analytic torsion of Ray and Singer 
\cite{Ray, RS1, RS_symp}. This numerical invariant 
is related to (but does not seem to coincide with) 
a quantity considered in \cite{Bismut_Lott}. Upon writing the semiclassical 
partition function in terms of a `twisted Ray-Singer metric', 
we use general results of \cite{Schwarz_resolvent, Adams_Sen}, 
in order to show that the extended Ray-Singer norm 
is independent of the metric data employed in the gauge-fixing 
procedure.
We also discuss the argument of the semiclassical partition function, 
which displays a metric dependence reminiscent of that 
known from usual Chern-Simons theory. 
Section 4 considers the generalized 
torsion of  graded D-brane pairs
(of unit relative grade) in the presence of a scalar condensate. 
For the case of trivial underlying flat connections, we show that 
the generalized torsion reduces to a power of the classical 
Ray-Singer invariant of the three-cycle. 

\

\noindent{\bf Conventions} There exist a few different conventions 
for the Ray-Singer torsion in the literature. In this note, we
define the Ray-Singer torsion
of a closed three-manifold $L$ (with respect to the trivial flat line bundle) 
by:
\be
\label{conventions}
T(L)=\prod_q{det^{',reg}(\Delta_q)^{\frac{q(-1)^q}{2}}}~~,
\ee
where $\Delta_q$ is the $q$-form Laplacian and $det^{',reg}(\Delta_q)$ is 
the regularized determinant of the restriction to the orthocomplement of its 
kernel.
Other papers define the Ray-Singer torsion to be $T(L)^{-1}, T(L)^{2}$ or 
$T(L)^{-2}$, where $T(L)$ is the quantity in (\ref{conventions}).
Our conventions for the generalized torsion will be an extension 
of (\ref{conventions}).

\section{Graded Chern-Simons theories and topological D-branes}

\subsection{The string field theory of graded topological A-branes}

Consider a special Lagrangian 3-cycle $L$ in a a Calabi-Yau threefold $X$, 
endowed with the `fundamental' orientation discussed in \cite{sc, gauge}. 
Given a collection of graded topological branes 
(of different grades $n$) wrapping $L$, 
we form the total bundle ${\bf E}=\oplus_{n}{E_n}$, endowed with the 
$\Z$-grading induced by $n$. Here $E_n$ are flat bundles which describe the 
worldvolume backgrounds of the various topological D-branes. 
Throughout this paper, we shall assume that the system contains a finite 
number of graded D-branes, i.e. $n$ takes values in a finite set of integers.
We also make the convention that a form on $L$ of rank lying outside 
the interval $0..3$ is defined to be zero. Note that we consider 
{\em complex} flat vector bundles $E_n$, which are not required to be unitary 
(i.e. there need not exist metrics on $E_n$ which are covariantly-constant 
with respect to the flat connections).

The graded Chern-Simons theory of \cite{sc, bv} describes sections  
$u$ of the bundle:
\be
{\cal V}=\Lambda^*(T^*L)\otimes End({\bf E})~~,~~
\ee
which we endow with the total grading 
${\cal V}=\oplus_{t}{{\cal V}^t}$, where:
\be
{\cal V}^t=\oplus_{\tiny \begin{array}{c}k, m, n\\k+n-m=t\end{array}}
{\Lambda^k(T^*L)\otimes Hom(E_m, E_n)}~~.
\ee
The degree of $u$ with respect to this grading is:
\be
|u|=rk u + \Delta(u)~~,
\ee
where $\Delta(u)=n-m$ if $u\in \Omega^*(L, Hom(E_m,E_n))$. The grading $|.|$ 
is related (after localization) 
to the worldsheet $U(1)$ charge of topological string states.
The space ${\cal H}=\Gamma({\cal V})$ of sections of ${\cal V}$ is
{\em total boundary space} of \cite{sc}, and can be viewed as the 
collection of open string states of the system. 
It is endowed with the grading ${\cal H}^k=\Gamma({\cal V}^k)$.

Triple string interactions are described by the so-called 
{\em total boundary product}, which is defined through:
\be
\label{bullet}
u\bullet v= (-1)^{\Delta(u)rk v}u\wedge v~~,
\ee
where the wedge product on the right hand side includes composition 
of morphisms in $End({\bf E})$. This associative product is compatible 
with the grading and admits the identity endomorphism of 
${\bf E}$ as a neutral element:
\be
\label{bullet_props}
|u\bullet v|=|u|+|v|~~, ~~1\bullet u=u\bullet 1=u~~.
\ee
(note that $|1|=0$). 

The direct sum $A^{(0)}=\oplus_n{A_n}$ of the flat connections on $E_n$
induces a flat structure on $End({\bf E})$. If $d^{(0)}$ 
is de Rham differential twisted by this flat connection,  
then it acts as a degree one derivation of the boundary product
(since the connection induced on $End({\bf E})$ 
is in the `adjoint representation').
To allow for more general backgrounds, we shift by elements 
$\phi\in {\cal H}^1$, which allows us to build the object 
$d=d^{(0)}+[\phi,.]_\bullet$, where $[.,.]_\bullet$ stands for the graded 
commutator\footnote{This is given by
$[u,v]_\bullet:=u\bullet v -(-1)^{|u||v|}v\bullet u$.}
 in the associative algebra $({\cal H}, \bullet)$. 
It is clear that $d$ is a degree one derivation of $({\cal H},\bullet)$
(since so are both terms in its definition): 
\be
\label{d_props}
|du|=|u|+1~~,~~d(u\bullet v)=(du)\bullet v+(-1)^{|u|}u\bullet (dv)~~.
\ee
In the language of 
\cite{Bismut_Lott}, $d$ is a `graded superconnection of total degree one'.
We refer the reader to \cite{Quillen} for basic facts regarding 
superconnections. 
For what follows, we shall pick a reference {\em flat}\footnote{Flatness
means that $d^2=0$, and will be required by the equations of motion.} 
graded superconnection $d$ (of degree one)
which need not coincide with the original flat connection $A^{(0)}$; 
the formalism is independent of this choice.

One also has a graded trace on $\Omega^*(L, End({\bf E}))$, which is defined through:
\be
\label{str}
str(u)=\sum_{n}{(-1)^n tr(u_{nn})}~~,~~{\rm for}~~u=\oplus_{m,n}{u_{mn}}~~,
\ee
with $u_{mn}\in \Omega^*(L, Hom(E_m, E_n))$. 
The bilinear form:
\be
\label{bf_A}
\langle u, v \rangle:=\int_{L}{str(u\bullet v)}~~ 
\ee
is non-degenerate and has the properties:
\bea
\label{form_invariance}
\langle u, v\rangle =(-1)^{|u||v|}\langle v, u \rangle~~,~~
\langle du, v\rangle +(-1)^{|u|}\langle u, dv\rangle=0~~,~~
\langle u\bullet v, w\rangle =\langle u, v\bullet w\rangle~~ 
\eea
It also obeys the selection rule $\langle u, v\rangle=0$ unless $|u|+|v|=3$,
so that the sign prefactor 
in the first equation of (\ref{form_invariance}) is always 
one (though it is more convenient to write it out explicitly, as we did).

The string field theory of \cite{sc,bv} is described by the action
\footnote{This is the real part of the complex action considered in 
\cite{gauge}. In that paper, we were interested in the moduli space, 
which can be understood without worrying about the real and imaginary parts.
In the present note, we shall consider the path integral, which requires that 
we work with the physical action, an object which must always be real.}:
\be
\label{action}
S(\phi)=\int_{L}{str\left[\frac{1}{2}\phi\bullet d\phi+\frac{1}{3}
    \phi\bullet\phi\bullet\phi\right]} +cc=
\frac{1}{2}\langle \phi, d\phi\rangle +\frac{1}{3}\langle
\phi,\phi\bullet \phi\rangle +cc~~~~,  
\ee which is defined on the degree one
component 
\be
{\cal H}^1=\{\phi\in {\cal H}||\phi|=1\}=
\Gamma(\oplus_{k+n-m=1}{\Lambda^k(T^*L)\otimes Hom(E_m,E_n)})
\ee 
of the total boundary space. 
This defines a `graded Chern-Simons field theory', which governs the 
dynamics of background shifts $\phi$. The equations of motion:
\be
\label{mc}
d\phi+\frac{1}{2}[\phi,\phi]_\bullet =0 \Leftrightarrow d\phi+
\phi\bullet \phi=0~~(\phi\in {\cal H}^1)~~
\ee
are equivalent with the requirement that the shifted 
superconnection $d_\phi=d+[\phi,.]_\bullet$ satisfies $(d_\phi)^2=0$. 
Hence the extrema of (\ref{action}) describe flat superconnections 
of total degree one. 

The equations of motion (\ref{mc}) are invariant
\footnote{One has 
$d(\phi^g)+\phi^g\bullet \phi^g=g\bullet(d\phi+\phi\bullet \phi)\bullet 
g^{-1}$. The object $d\phi+\phi\bullet \phi$ is the curvature of the 
associated superconnection.} 
with respect to gauge transformations of the form: 
\be
\label{Gauge}
\phi\rightarrow \phi^g=g\bullet \phi \bullet g^{-1}+g\bullet dg^{-1}~~,
\ee
where $g$ is an invertible element of the subalgebra $({\cal H}^0, \bullet)$.
This means that $g$ belongs to the group of units:
\be
\label{Gauge_group}
{\cal G}=\{g\in {\cal H}^0|{\rm~exists~}~g^{-1}\in {\cal H}^0~{\rm~such~that~}
~g\bullet g^{-1}=g^{-1}\bullet g=1\}~~.
\ee
One has the standard transformation rule for the `covariant differential'
$d_\phi$:
\be
d_\phi(g^{-1}\bullet u\bullet g)=g^{-1}\bullet d_{\phi^g}\bullet g
\Leftrightarrow d_\phi(Ad_{g^-1}u)=Ad_{g^{-1}}(d_{\phi^g}u)~~,
\ee
for $u\in {\cal H}$, where $Ad_{g}(u):=g\bullet u \bullet g^{-1}$.
The moduli space results upon dividing the space ${\cal S}$ of solutions to 
(\ref{mc}) through the gauge symmetries (\ref{Gauge}). 
This is a standard Maurer-Cartan problem.

In general, the 
the action (\ref{action}) is invariant only under `small' gauge 
transformations \cite{gauge}, i.e. those transformations for which 
$g$ can be written in the form:
\be
\label{exp}
g=e^\alpha_\bullet :=\sum_{k\geq 0}{\frac{1}{k!}\alpha^{\bullet k}}~~,
\ee
where $\alpha^{\bullet k}$ stands for $k$-th iteration of 
the $\bullet$-product of $\alpha$ with itself (and we define 
$\alpha^{\bullet 0}:=1$).

For infinitesimal $\alpha$, the gauge 
transformations (\ref{Gauge}) become:
\be
\label{gauge}
\phi\rightarrow\phi+\delta_\alpha\phi~~, 
\ee 
with $\delta_\alpha\phi=-d\alpha-[\phi,\alpha]_\bullet$. 
For small $\phi$ and $\alpha$, 
the moduli problem  reduces to its linearized version:
\be
\label{mc_lin}
d\phi=0~~,~~\phi\equiv \phi-d\alpha~~,
\ee
which describes infinitesimal deformations of the background.

\paragraph{Observation} Since graded Chern-Simons theories contain higher rank 
forms, a complete analysis of their gauge symmetries (with a view toward 
quantization) requires the Batalin-Vilkovisky formalism. The classical part 
of this analysis was carried out in \cite{bv}, while gauge fixing and 
propagators are discussed in \cite{gbv}. In the present paper, we 
will be able to avoid the BV formalism by employing the 
`resolvent method' of \cite{Schwarz_resolvent, Schwarz_resolvent2}. 
The BV analysis gives the same results (as expected 
from the general observations of \cite{Blau_Thompson}). 

\subsection{D-brane interpretation of the background}

Our string field theory is formulated around a specific 
background described by certain collections of graded D-branes. 
A shift of the background (which can be achieved by use of a solution $\phi$ 
to (\ref{mc})) deforms the differential $d$ into $d_\phi u=du+[\phi,u]$. 
It is easy to check (see \cite{com1, com3, gauge}) 
that such shifts preserve the relevant  properties of $d$. 
When expanded around the shifted background, the 
string field action has the same form (\ref{action}) up to the addition 
of an irrelevant constant \cite{com1}.  

As explained in \cite{com1, com3}, a general background admits a D-brane 
interpretation which allows for a systematic description in the language of 
category theory. This is related to the basic fact \cite{Bismut_Lott} 
that degree one superconnections can be decomposed as the sum of 
a usual connection and a collection of bundle-valued forms, which can be 
interpreted as condensates of spacetime fields associated with 
boundary condition changing operators. The decomposition properties of 
the boundary product and bilinear form with respect to these fields
can then be used to extract the D-brane content of the background
\cite{com1, com3}. The D-brane interpretation is different at various 
points in the moduli space. This identifies our moduli space of vacua 
with the moduli space of topological D-brane composites \cite{com1, com3}.
We refer the reader to the papers just cited for a detailed 
presentation of this analysis. It was shown on general grounds in \cite{com1}
(and, for the models discussed here, in \cite{sc}) that the collection of 
D-brane composites describing a background $\phi$ defines a `differential 
graded category', part of 
which can be related to certain enhanced triangulated 
categories \cite{BK}\footnote{We warn the reader that the description 
of `D-brane category dynamics' in \cite{com1,com3} is only local, 
i.e. based on the linearized version (\ref{mc_lin}) of the equations 
of motion; similar limitations apply to the triangulated category picture, 
which is a consequence of this. This is related to 
the issue of finding a good global definition for the 
`moduli space of a triangulated category', which seems to be best approached 
at the string field theory level.}. For the B-model
version of the theory (which was considered in \cite{Diaconescu}), 
the relevant triangulated category essentially coincides with 
the derived category of coherent sheaves. 

From many points of view, the string field theory description must come 
{\em before} any discussion in terms of triangulated categories. For example, 
a proper definition of deformations seems to require 
string field data, and possesses a standard (i.e. Maurer-Cartan) 
formulation only at the string field level \cite{gauge}.

\section{The semiclassical approximation and generalized Ray-Singer torsion}

We are interested in the partition function $Z$ of (\ref{action}). 
Fixing a flat superconnection $A$, this can be determined by performing a 
background perturbation expansion around $A$. It is a standard 
fact that the semiclassical contribution to this expansion is given (up to the 
trivial factor $e^{-i\lambda S(A)}$, which we shall ignore) by certain
functional determinants, which encode the contribution of the kinetic term.
The purpose of this section is to evaluate this approximation in order 
to extract a certain numerical invariant associated 
with our models. 

\subsection{Hermitian structure}

Our gauge-fixing procedure will require that we pick a Riemannian 
metric $g$ on $L$ and Hermitian metrics on the bundles
$E_n$. The last determine a Hermitian metric $g_{\bf E}$ on ${\bf E}$, 
and thus a Hermitian metric on $End({\bf E})$, given by:
\be
(\alpha,\beta)=tr(\alpha^\dagger\circ \beta)~~{\rm~for~}~~\alpha,\beta\in 
End({\bf E}_p)~~,p\in L~~,
\ee
where $\alpha^\dagger$ is the Hermitian conjugate of $\alpha$ with respect to 
$g_{\bf E}$. On the other hand, $g$ induces a Hermitian 
metric $(.,.)$ on $\Lambda^*(T^*L)$
in the standard fashion. This is given by: 
\be
(*{\overline \omega})\wedge \eta=(\omega, \eta)vol_g~~,{\rm~for~}~~
\omega, \eta\in \Lambda^*(T^*_pL)~~,
\ee
where $vol_g$ is the volume form induced by $g$ on $L$ (with respect to the 
fundamental orientation on $L$), while 
$*$ is the {\em complex linear} Hodge operator, 
which is involutive and satisfies:
\be
rk(*\omega)=3-rk\omega~~.
\ee
Combining these, we obtain a Hermitian metric $(.,.)_{\cal V}$ on the bundle 
${\cal V}=\Lambda^*(T^*L)\otimes End({\bf E})$, which can be described through
the relation:
\be
\label{Vmetric}
tr(*u^\dagger\wedge v)=(u,v)_{\cal V}
vol_g~~{\rm~for~}u,v\in {\cal V}_p~~,~~p\in L~~.
\ee
It is clear that $(u,v)_{\cal V}$ vanishes (on bi-homogeneous elements) 
unless $\Delta(u)=\Delta(v)$ and $rk u= rk v$ (and thus unless $|u|=|v|$). 
In particular, we have induced scalar products on each of the subbundles 
${\cal V}^k$, which are mutually orthogonal with respect to (\ref{Vmetric}).
Integration over $L$ gives a Hermitian scalar product on the 
space ${\cal H}=\Gamma({\cal V})$:
\be
\label{hmetric}
h(u,v)=\int_{L}{(u,v)vol_g}=
\int_{L}{tr(*u^\dagger\wedge v)}~~,~~{\rm for}~~u,v\in {\cal H}~~.
\ee

\subsubsection{The conjugation operator} 

This was discussed in detail \cite{gauge}, so I will be short. 
Since the bilinear form (\ref{bf_A}) is non-degenerate, there exists a unique 
antilinear operator $c$ on ${\cal H}$ with the property:
\be
\label{h}
h(u,v)=\langle cu,v\rangle=\int_{L}{str[(cu)\bullet v]}~~.
\ee
This has the following form on decomposable elements:
\be
\label{c_A}
c(\omega\otimes f)=(-1)^{n+\Delta(f)(1+rk \omega)}
(*{\overline \omega})\otimes f^\dagger~~,~~{\rm~for~}~~\omega\in \Omega^*(L)~~
{\rm~and~}~~f\in Hom(E_m,E_n)~~.
\ee
The conjugation operator satisfies:
\be
\label{c_props}
|cu|=3-|u|~~, ~~c^2=id~~,
\ee
and thus obeys the axiomatic framework of \cite{superpot}.
Note that $c$ is anti-unitary:
\be
h(u,v)=h(cv,cu)~~.
\ee
In the ungraded case (${\bf E}=E_0$), 
$c$ reduces to the {\em antilinear} Hodge operator 
${\overline *}$, coupled to the bundle $End(E)$.

\subsubsection{The adjoint of $d$ and the Laplacian}

As mentioned above, the 
scalar product $h$ satisfies the selection rule:
\be
\label{h_sel}
h(u,v)=0 ~~,~~{\rm~unless~}~~|u|=|v|~~.
\ee
Considering the Hermitian conjugate $d^\dagger$ of $d$ (with respect to $h$), we have:
\be
d^\dagger u= (-1)^{|u|}cdcu~~,~~|d^\dagger u|=|u|-1~~,~~(d^\dagger)^2=0~~.
\ee
Using this operator, one constructs the `deformed Laplacian' 
$\Delta=dd^\dagger+d^\dagger d$, which is an order two elliptic operator
on ${\cal H}$. We also note the relations:
\bea
\label{cdd}
cd^\dagger u= (-1)^{|u|}dcu~~&,&~~d^\dagger cu=(-1)^{|u|+1}cdu~~\nn\\
d^\dagger dc=cdd^\dagger~~&,&~~dd^\dagger c=cd^\dagger d~~.
\eea

\subsection{The spacetime ghost grading}

According to (\ref{h_sel}), we have induced 
Hermitian scalar products on each of 
the subspaces ${\cal H}^k$, which are mutually orthogonal with respect 
to $h$. For what follows, it will be convenient to use the grading 
$s(u)=1-|u|$, which in the BV formalism corresponds to the ghost number 
of the string field theory \cite{bv}. 
We shall use the notation ${\cal H}(\sigma)={\cal H}^{1-\sigma}$ 
for the homogeneous subspaces of ${\cal H}$ with respect to this grading, 
and the notation $H_\sigma({\cal H})=H^{1-\sigma}({\cal H})$ for the 
associated components of $H^*({\cal H})$ (since the ghost grading 
is decreased by $d$, this takes the original cochain complex $({\cal H},d)$ 
into a chain complex, which is why we use homological notation).
With this convention, the string field lies in the subspace ${\cal H}(0)$ 
of vanishing ghost number. 
Note that:
\be
\label{s_props}
s(du)=s(u)-1~~,~~s(cu)=-1-s(u)~~,~~s(cdu)=-s(u)~~.
\ee
In particular, $cd$ induces a well-defined antilinear operator on the 
physical subspace ${\cal H}(0)$. 

It is clear from the definition of $c$ that a nonzero element $u\in {\cal H}$ 
always has a nonzero conjugate partner $cu$. Since the number of graded 
components $E_n$ is assumed to be finite, there exists 
a maximal value of the degree $s$, which we shall denote by $N$.
The second relation in (\ref{s_props}) then shows that the minimal value of 
$s$ is $-1-N$. Accordingly, the degree $|.|=1-s(.)$ lies in the interval
$1-N\dots 2+N$. 

We also recall the standard decompositions:
\be
{\cal H}=K\oplus im d \oplus im d^\dagger~~,~~ker d=imd \oplus K~~,~~ker d^\dagger= im d^\dagger \oplus K~~,
\ee
where $K=ker \Delta=ker d\cap ker d^\dagger$. Hodge theory leads 
to the identification $K\approx H_d^*({\cal H})$.

\subsection{Complex and real determinants and induced Euclidean scalar 
products}

In the next subsection, it will be convenient to `decompose our fields
into real components' when performing the path integral. For that purpose, 
we collect some useful facts and definitions. This is entirely trivial and 
well-known, but I shall state it explicitly nonetheless. 

The complex vector space ${\cal H}$ 
can be viewed as a real vector space upon restriction 
of the field of scalars. Then the Hermitian product $h$
induces an Euclidean scalar product $(.,.):=Re h$ on ${\cal H}$. 
Given a complex vector subspace $W$ of ${\cal H}$, it is easy to check 
that for any vector $u$ in ${\cal H}$, one has the equivalence:
\be
\label{EH}
(u,v)=0{\rm ~for~all~}v\in W\Leftrightarrow <u,v>=0{\rm~for~all~}v\in W.
\ee
Indeed, if $W$ contains a vector $v$, then it also contains the vector 
$iv$, and 
one has $(u,iv)=Reh(u,iv)=Re [ih(u,v)]= -Im h(u,v)$. Relation (\ref{EH}) 
shows that the orthogonal complements of $W$ with respect to $h$ and 
$(.,.)$ coincide, so we shall use the symbol $W^\perp$ to denote either.

A complex-linear operator $O$ on a complex vector space $V$ 
can also be viewed as a real-linear operator on the underlying 
real vector space. 
Therefore, one has two notions for the determinant of $O$, namely the
complex and real determinant.
The first is the determinant of $O$ viewed as a complex-linear map, 
i.e. the determinant of its matrix in a complex-linear basis
$e_\alpha$ of $V$:
\be
\label{real_matrix}
Oe_\alpha=a_{\beta\alpha}e_\beta\Rightarrow det O=det (a_{\alpha\beta})~~.
\ee
It can be described more geometrically by considering the maximal exterior 
power $\Lambda^d V$ of $V$ viewed as a complex vector space
(here $d$ is the complex dimension of $V$). Indeed, $O$ induces a linear 
endomorphism of $\Lambda^d V$. Since the latter space is complex 
one-dimensional, this endomorphism is canonically identified with a complex 
number, which is the determinant of $O$. In fact, $\Lambda^d V$
has a basis given by $e_1\wedge\dots \wedge e_d$, and 
$O(e_1)\wedge\dots \wedge O(e_d)=det O e_1\wedge\dots \wedge e_d$.

The real determinant arises from a similar construction, where one views 
$O$ as a real linear operator on the real vector space $V$ (whose 
real dimension is $2d$). The latter admits the basis 
$e_1\dots e_d, ie_1\dots ie_d$, which gives:
\bea
Oe_\alpha&=&x_{\beta\alpha}e_\alpha+y_{\beta\alpha}(ie_\alpha)~~\nn\\
O(ie_\alpha)&=&-y_{\beta\alpha}e_\alpha+x_{\beta\alpha}(ie_\alpha)~~,
\eea
where $x_{\alpha\beta}=Re a_{\alpha\beta}$ and 
$y_{\alpha\beta}=Im a_{\alpha\beta}$. The real determinant $det_\R O$ 
is the determinant of the resulting real $2d\times 2d$ matrix. This can again be 
identified with the operator induced by $O$ on the real one-dimensional 
vector space $\Lambda_\R^{2d} V$, where $\Lambda_\R$ denotes the real 
exterior product. One has $O(e_1)\wedge_\R\dots \wedge_\R O(e_d)
\wedge_\R O(ie_1)\wedge_\R \dots \wedge_\R O(ie_d)=det_R(O) 
e_1\wedge_\R \dots \wedge_\R e_d\wedge (ie_1)\wedge_\R \dots \wedge_\R(ie_d)$.
One can show that:
\be
det_\R O= |det O|^2~~,
\ee
a formula which will be used repeatedly in later subsections.

\subsection{The semiclassical partition function in the 
resolvent formalism}

\subsubsection{The computation} 

To compute the semiclassical partition function, 
we use the resolvent formalism of \cite{Schwarz_resolvent,Schwarz_resolvent2}, 
in its generalized version discussed in \cite{Adams_Sen0, Adams_Sen}\footnote{
A streamlined presentation of this material can be found in the thesis of 
D. H. Adams. The author thanks him for bringing this thesis to his attention.}
(see \cite{Adams_scl, Adams_FP, Adams_cpct} and 
\cite{GK, NC} for related issues). For this, we focus on the kinetic term: 
\be
\label{Skin}
S_{kin}(\phi)=Re \langle \phi, d\phi\rangle=
Re \int_{L}{str(\phi \bullet d\phi)}~~,
\ee
where $d:=d_A$ is the differential on ${\cal H}$ associated with the
background $A$. The semiclassical partition function is formally given by:
\be
\label{Zformal}
Z_{scl}=\int{{\cal D}[\phi]e^{-i\lambda S_{kin}(\phi)}}~~,
\ee
where $\lambda>0$ is the coupling constant.
This path integral is of course ill-defined. The most severe problem 
is the presence of zero modes, which are related to the gauge invariance 
$\phi\rightarrow \phi+d\omega$, with $\omega$ an element of 
${\cal H}(1)={\cal H}^0=
\oplus_{m,n}{\Omega^{m-n}(L, Hom(E_m,E_n))}$. A precise 
characterization of zero modes is as follows. Using the metric 
induced on ${\cal H}(0)$, we write $S_{kin}$ in the form:
\be
S_{kin}(\phi)=Re~h(\phi, cd\phi)=(\phi, cd\phi)~~,
\ee
where $(.,):=Reh(.,.)$ is the Euclidean scalar product induced by 
$h$, if we view ${\cal H}^1={\cal H}(0)$ as a real vector 
space. 
The zero modes lie in the kernel of this quadratic functional.
Since $(.,.)$ is nondegenerate, this coincides with the kernel of the 
antilinear operator $T_0:=cd:{\cal H}(0)\rightarrow {\cal H}(0)$. 
In fact, $T_0$ becomes a (real-linear) 
self-adjoint operator, as can be checked by using invariance of 
$\langle.,.\rangle$ with respect to $d$:
\be
(T_0u, v)=(u,T_0v)~~{~~\rm~for~~}~u, v \in {\cal H}(0)\Rightarrow 
T_0^t=T_0~~,
\ee
where $^t$ stands for the adjoint with respect to $(.,.)$.
It is also easy to check that:
\be
(\phi, T_0\phi)=(\phi_M, T_0\phi_M)~~,
\ee 
where we decomposed $\phi:=\phi_M\oplus \phi_K$ with 
$\phi_K\in ker d=ker T_0$ and $\phi_M \in (ker d)^\perp=(ker T_0)^\perp$. 
If we let $T_0'$ denote the restriction of $T_0$ to the orthogonal 
complement of its kernel, then formal integration in (\ref{Zformal}) leads to:
\be
\label{Zformal1}
Z_{scl}=vol(ker(T_0))\int{{\cal D}[\phi_M]e^{-i\lambda (\phi_M, T_0'\phi_M)}}~~.
\ee

Both factors in (\ref{Zformal1}) are ill-defined, 
since the first involves the `volume' of a vector space
while the second is a Gaussian integral for a generally indefinite 
quadratic form. The second problem is solved in standard manner by 
replacing $T_0 '$ with the strictly positive operator 
$|T_0'|=\sqrt{T_0'T_0'}=
\sqrt{(d^\dagger d)'_{s=0}}$, 
where the subscript $s=0$ indicates that 
$d^\dagger d$ is restricted to the subspace ${\cal H}(0)$.
This leads to the expression:
\be
\int{{\cal D}[\phi_M]e^{-i\lambda (\phi_M, T_0'\phi_M)}}=
ct\times (det_\R|T_0'|)^{-1/2}~~
\ee
where $det_\R|T_0'|$ denotes the real determinant of $|T_0'|$. 
Using $det_\R |T'_0|^{-1/2}=det |T_0'|^{-1}= (det'_{s=0}(d^\dagger d))^{-1/2}$,
we obtain:
\be
\label{Zformal2}
Z_{scl}=ct\times vol(ker(T_0)) (det'_{s=0}d^\dagger d)^{-1/2}~~,
\ee
where, for a Hermitian operator $O$, 
$det' O$ denotes the complex determinant of $O|_{(ker O)^\perp}$.

The last factor of (\ref{Zformal2}) is still ill-defined, since it involves 
an infinite product of eigenvalues. This is a standard problem, which 
is solved by zeta-function regularization.
We remind the reader that, given a positive elliptic operator $O$, 
acting on sections of a vector bundle over $L$, 
its zeta function $\zeta_O(z)$ $(z\in \C)$ is defined as follows.
For sufficiently large $Rez$, $\zeta_O(z)$ is given by the expansion:
\be
\zeta_O(z)=\sum_{\lambda}{\frac{n_\lambda}{\lambda^z}}~~,
\ee
where the sum is over the (strictly positive) 
eigenvalues $\lambda$ of $O$, whose multiplicities we denote by 
$n_\lambda$. 
The analytic continuation of this series to the 
complex plane is meromorphic and regular at the origin. 
This allows one to define the regularized determinant through the expression:
$det^{reg}(O):= e^{-\zeta'_{O}(0)}$, where 
$\zeta'_{O}(z):=\frac{d}{dz}\zeta_{O}(z)$. 
Applying this to the operator $|T_0'|^2$, we obtain:
\be
\label{detprime}
det'^{,reg}_{s=0}(d^\dagger d)= e^{-\zeta'_{(d^\dagger d)'_{s=0}}(0)}~~.
\ee

To regularize the volume factor in (\ref{Zformal2}), 
we use the so-called {\em method of resolvents} 
\cite{Schwarz_resolvent, Adams_Sen0, Adams_Sen},
which proceeds as follows\footnote{Since we work with complex fields, 
(i.e. $\phi$ varies in a complex vector space)
we will have to make some minor modifications in order to adapt the work of 
\cite{Adams_Sen}.}. First, we notice that the kinetic 
operator $T_0=cd$ admits an {\em elliptic resolvent}.
More precisely, one has the complex of {\em real} vector spaces:
\be
\label{res}
({\cal R})~~:~~0\rightarrow{\cal H}(N)
\stackrel{T_N=d}{\longrightarrow}\dots 
\stackrel{T_2=d}{\longrightarrow} 
{\cal H}(1)\stackrel{T_1=d}{\longrightarrow}{\cal H}(0) 
\stackrel{T_0=cd}{\longrightarrow}{\cal H}(0)~~,
\ee
(where $N$ denotes the maximal value of the spacetime ghost degree, 
as discussed in Subsection 3.2), and the associated Laplace operators:
\be
\label{Laplacians}
\Delta_\sigma:=T_\sigma^t T_\sigma+
T_{\sigma+1} T_{\sigma+1}^t
= d^\dagger d+dd^\dagger:{\cal H}(\sigma)\rightarrow {\cal H}(\sigma)
\ee
(with $T_{N+1}:=0$) are elliptic for all $\sigma=0\dots N$. 
To arrive at (\ref{Laplacians}), 
we used the fact that $T_\sigma^t=T_\sigma^\dagger$ for 
$\sigma>0$ and $T_0^t T_0=d^\dagger d$ for $\sigma=0$.
Notice that ${\cal H}(\sigma)={\cal H}^{1-\sigma}=\Gamma({\cal V}^{1-\sigma})$
are in fact complex vector spaces, but the last map in (\ref{res}) 
is complex antilinear, rather than complex linear.  This is why we 
can only view (\ref{res}) as a complex of real-linear maps. 

To regularize 
$vol(ker T_0)$, one picks auxiliary Hermitian metrics 
$h_\sigma^H$ on the homology spaces 
$H_\sigma({\cal R})=ker T_\sigma/im T_{\sigma+1}=H^{1-\sigma}_d({\cal H})$
of the resolvent, and considers the Hodge isomorphisms:
\be
\label{fsigma}
f_\sigma:K(\sigma)=K^{1-\sigma}=
ker T_\sigma\cap ker T_\sigma^t
\stackrel{\approx}{\longrightarrow} 
H_\sigma({\cal R})=H^{1-\sigma}_d({\cal H})~~
\ee 
induced by the obvious 
projections $ker T_\sigma\rightarrow H_\sigma({\cal R})$. 
Given this data, one can formally compute \cite{Adams_Sen}:
\bea
\label{vol_formal0}
vol(ker T_0)=\prod_{\sigma=0}^N{
\left[\left(\frac{det'_\R(T_{\sigma+1}^t T_{\sigma+1})}
{det_\R(f_\sigma^t f_\sigma)}\right)^{1/2}vol(H_\sigma({\cal R}))
vol({\cal H}(\sigma+1))\right]^{(-1)^{\sigma}}}~~,
\eea
where $det_\R$ denotes the determinant of real-linear maps and
we defined $det'_\R(T_{N+1}^t T_{N+1}):=1$ and $vol({\cal H}(N+1)):=1$. 
Relation (\ref{vol_formal0}) gives:
\bea
\label{vol_formal}
vol(ker T_0)=\prod_{\sigma=0}^N{
\left[\frac{det'_{s=\sigma+1}(d^\dagger d)}
{det(f_\sigma^\dagger f_\sigma)}vol(H_\sigma({\cal R}))vol({\cal H}(\sigma+1))
\right]^{(-1)^{\sigma}}}~~
\eea
(where again we define $det'_{s=N+1}(d^\dagger d):=1$).
This expression is given a meaning by dropping 
the ill-defined factors
$vol(H_\sigma({\cal R}))$ and $vol({\cal H}(\sigma))$ 
and using zeta-function regularization for
the determinants of the positive operators 
$(d^\dagger d)'_{s=\sigma}=
[(d^\dagger d)|_{ker (d)^\perp}]_{s=\sigma}$: 
\be
vol^{reg}(ker T_0)=\prod_{\sigma=0}^N{
\left[\frac{det^{',reg}_{s=\sigma+1}(d^\dagger d)}
{det(f_\sigma^\dagger f_\sigma)}\right]^{(-1)^{\sigma}}}~~.
\ee
Combining with (\ref{Zformal2}), we conclude:
\bea
\label{Zf}
Z_{scl}=Cdet'^{,reg}_{s=0}(d^\dagger d)^{-1/2}
\prod_{\sigma=0}^N{
\left[\frac{det'^{,reg}_{s=\sigma+1}(d^\dagger d)}
{det(f_\sigma^\dagger f_\sigma)}\right]^{(-1)^{\sigma}}}=\\
C e^{\frac{1}{2}\zeta'_{(d^\dagger d)'_{s=0}}(0)
+\sum_{\sigma=1}^N{(-1)^\sigma \zeta'_{(d^\dagger d)'_{s=\sigma}}(0)}}
\prod_{\sigma=0}^N{
\left[det(f_\sigma^\dagger f_\sigma)^{(-1)^{\sigma+1}}\right]}~~,\nn
\eea
where $C$ is a complex constant. If one uses the normalization conventions
of \cite{Adams_Sen}, then $C$ can be written as:
\be
\label{C}
C=\left(\frac{\pi}{\lambda}\right)^{\zeta_{|T_0|}(0)/2}e^{-\frac{i\pi}{4}
\eta_{T_0}(0)}~~,
\ee
where $\zeta_{|T_0|}:=\zeta_{|T_0'|}$
is the zeta-function of the positive operator $|T_0'|$ and 
$\eta_{T_0}:=\eta_{cd_{s=0}}$ is the eta-function of $T_0$. 
The latter is defined through analytic continuation of 
the following expression\footnote{Remember that $T_0$ is 
self-adjoint only as a real-linear operator. To find its eigenvalues, 
one must of course consider the complexification $W$ of the underlying
real vector space of ${\cal H}(0)$. The complex dimension of $W$ 
equals $2dim_\C {\cal H}(0)$. 
The selfadjoint operator $T_0$ induces a 
{\em complex-linear} Hermitian operator $T_0^\C$
on $W$. It is the real 
eigenvalues of this operator which appear in (\ref{eta})  
for our case.}
(which is valid for $Re z>>0$):
\be
\label{eta}
\eta_{T_0}(z)=\sum_{\lambda>0}{\frac{n_\lambda}{\lambda^z}}-
\sum_{\lambda<0}{\frac{n_\lambda}{|\lambda|^z}}~~,
\ee
the sums being over the strictly positive and strictly negative 
eigenvalues of $T_0$ $(n_\lambda$ are the multiplicities). We refer the reader 
to \cite{Adams_Sen} for the justification of (\ref{C}) (in particular, 
it is shown there that $\eta_{T_0}(0)$ is well-defined). Here we only note 
that the absolute value of $C$ produced by a putative non-perturbative 
solution of the full graded Chern-Simons theory 
need not strictly agree with (\ref{C}), 
due to corrections from a possibly nontrivial gauge stabilizer
of the background superconnection $A$ (this is 
similar to what happens for the ungraded case 
\cite{Adams_Sen, Adams_FP, Adams_scl, Adams_cpct}).
Therefore, the modulus of expression (\ref{C}) should {\em not} 
be taken at face value.

\subsubsection{Generalized Ray-Singer torsion}

Let us write the partition function (\ref{Zf}) in an alternate form. 
First we introduce the notation 
$\zeta_{\sigma}(z):=\zeta_{(d^\dagger d)'_{s=\sigma}}(z)$.
Lemma 3.1 of \cite{Adams_Sen} establishes the relation:
\be
\label{zeta_rel}
\zeta_\sigma(z)+\zeta_{\sigma+1}(z)=\zeta_{\Delta_\sigma}(z)
~~(\sigma \geq 0)~~.
\ee
where $\zeta_{\Delta_\sigma}$ is a shorthand for $\zeta_{\Delta'_\sigma}$.
This follows from the fact that 
$\Delta'_\sigma=(dd^\dagger)'\oplus (d^\dagger d)'$ (since 
$(ker \Delta)^\perp=im d^\dagger \oplus im d=(ker d)^\perp\oplus 
(ker d^\dagger)^\perp=ker (d^\dagger d)^\perp\oplus (ker d d^\dagger)^\perp$),
which implies
$\zeta_{\Delta'_\sigma}(z)=\zeta_{(d^\dagger d)'_\sigma}(z)+
\zeta_{(dd^\dagger)'_\sigma}(z)$.
Then one notices that $d$ gives an isomorphism between 
$ker(d^\dagger d)^\perp=ker (d)^\perp$ and 
$ker(dd^\dagger)^\perp=ker(d^\dagger)^\perp=im(d)$, thereby inducing 
a bijection between the nonzero eigenvalues of $(d^\dagger d)_{\sigma+1}$ and 
$(dd^\dagger)_\sigma$ (if $d^\dagger du =\lambda u$ then 
$dd^\dagger du=\lambda du$ etc.). Therefore, 
$\zeta_{(dd^\dagger)'_\sigma}(z)=\zeta_{\sigma+1}(z)$, which leads to
(\ref{zeta_rel}).

Equation (\ref{zeta_rel}) implies:
\be
\label{s1}
\zeta_0+2\sum_{\sigma=1}^N{(-1)^\sigma\zeta_\sigma}=
\sum_{\sigma=0}^N{(-1)^\sigma(1+2\sigma)\zeta_{\Delta_\sigma}}~~.
\ee
The next step is to notice that the last two equations in 
(\ref{cdd}) imply $\Delta c=c\Delta$ and thus:
\be
c\Delta_\sigma=\Delta_{-1-\sigma}c~~,~~\Delta_\sigma c=c\Delta_{-1-\sigma}~~,
\ee
because $s(cu)=-1-s(u)$. 
Since $c$ is a bijection, this shows that the operators
$\Delta_\sigma$ and $\Delta_{-1-\sigma}$ are iso-spectral (i.e. have the 
same eigenvalues, with the same multiplicities). Therefore:
\be
\zeta_{\Delta_\sigma}(z)=\zeta_{\Delta_{-1-\sigma}}(z)~~{\rm~~for~~}
\sigma\in \{ -1-N \dots N\}~~.
\ee
Thus:
\be
\label{s2}
\sum_{\sigma=0}^N{(-1)^\sigma(1+2\sigma)\zeta_{\Delta_\sigma}}=
\sum_{\sigma=-N-1}^N{(-1)^\sigma\sigma\zeta_{\Delta_\sigma}}=
\sum_{q=1-N}^{N+2}{(-1)^q q\zeta_{\Delta^{(q)}}}~~,
\ee
where $\Delta^{(q)}=\Delta_{1-q}$ is the Laplacian on the space 
${\cal H}^q={\cal H}(1-q)$. 
Combining (\ref{s1}) and (\ref{s2}), we obtain:
\be
\label{sums_final}
\frac{1}{2}\zeta_0+\sum_{\sigma=1}^N{(-1)^\sigma\zeta_\sigma}=
\frac{1}{2}\sum_{\sigma=-N-1}^{N}{(-1)^\sigma\sigma\zeta_{\Delta_\sigma}}=
\frac{1}{2}\sum_{q=1-N}^{N+2}{(-1)^q q\zeta_{\Delta^{(q)}}}~~,
\ee
where $\sigma$ and $q$ in the right hand side 
run over their maximal domains of definition 
$\{-1-N\dots N\}$ and $\{1-N\dots N+2\}$. 
Let us define the quantity:
\be
\label{T}
T(L, A):=\prod_{q=1-N}^{N+2}{[det^{',reg}\Delta^{(q)}]^{\frac{(-1)^qq}{2}}}=
e^{-\frac{1}{2}\sum_{q=1-N}^{N+2}{(-1)^q q\zeta'_{\Delta^{(q)}}(0)}}=
e^{-\frac{1}{2}
\sum_{\sigma=-N-1}^N{(-1)^\sigma\sigma \zeta'_{\Delta_\sigma}(0)}}~~,
\ee
which is a generalized version of the analytic 
torsion of Ray and Singer 
\cite{Ray, RS1, RS_symp}. Then:
\be
Z_{scl}=C T(L, A)^{-1}
\prod_{\sigma\geq 0}{
\left[det(f_\sigma^\dagger f_\sigma)^{(-1)^{\sigma+1}}\right]}~~.
\ee
It is convenient to write $Z_{scl}=C{\tilde Z}_{scl}$, where:
\be
{\tilde Z}_{scl}=T(L, A)^{-1}
\prod_{\sigma\geq 0}{
\left[det(f_\sigma^\dagger f_\sigma)^{(-1)^{\sigma+1}}\right]}~~.
\ee

{\bf Observation} Relations (\ref{cdd}) also imply that the 
operators $(d^\dagger d)_\sigma|_{ker(d)^\perp}$ and 
$(dd^\dagger)_{-1-\sigma}|_{ker (d^\dagger)^\perp}$ are iso-spectral.
Since the latter is iso-spectral with 
$(d^\dagger d)_{-\sigma}|_{(ker d)^\perp}$, this gives the relation:
\be
\zeta_\sigma(z)=\zeta_{-\sigma}(z)~~,
\ee
which implies $\frac{1}{2}\zeta_0+\sum_{\sigma=1}^N{(-1)^\sigma\zeta_\sigma}=
\frac{1}{2}\sum_{\sigma=-N}^N{(-1)^\sigma\zeta_\sigma}$.
Together with (\ref{sums_final}), this gives:
\be
\frac{1}{2}\sum_{q=1-N}^{N+2}{(-1)^q q\zeta_{\Delta^{(q)}}}=
\frac{1}{2}\sum_{\sigma=-N}^N{(-1)^\sigma\zeta_\sigma}~~
\ee
and leads to the following expression for the torsion:
\be
T(L,A)=e^{-\frac{1}{2}
\sum_{\sigma=-N}^N{(-1)^\sigma\zeta'_{\sigma}(0)}}~~.
\ee
Note that the sum in the exponent does not contain a term with 
$\sigma=-N-1$.

\subsubsection{A Hermitian analogue of the Ray-Singer metric}

The factor $J:=\prod_{\sigma\geq 0}{
\left[det(f_\sigma^\dagger f_\sigma)^{(-1)^{\sigma+1}}\right]}$
can be described geometrically as follows. 
We first notice that there exists a unique antilinear involution
$c_*:H_{\sigma}({\cal R})\rightarrow H_{-1-\sigma}({\cal R})$ with the 
property:
\be
f_\sigma c=c_*f_{-1-\sigma}~~,
\ee
where we consider the Hodge isomorphisms 
(\ref{fsigma}) for all $\sigma=-N-1\dots N$. Given this involution, we 
use the metrics $h^H_\sigma$ for $\sigma \geq 0$ to introduce
metrics on the spaces $H_\sigma({\cal R})=
H^{1-\sigma}({\cal H})$ with $\sigma<0$ through the relations:
\be
h^H_\sigma(\alpha,\beta)=h^H_{-1-\sigma}(c_*\beta,c_*\alpha)
~~{\rm~for~all~}~~\sigma=-N-1 \dots -1~~.
\ee
With respect to these metrics, the operators $c_*$ are anti-unitary.
Use of particular extension of the auxiliary data $h^H_\sigma (\sigma\geq 0)$ 
to the entire cohomology $H^*_d({\cal H})$ is crucial for the validity 
of certain arguments below. 

We have $f_\sigma^\dagger f_\sigma=cf_{-1-\sigma}^\dagger f_{-1-\sigma} c$, 
which implies:
\be
det(f^\dagger_\sigma f_\sigma)=det (f^\dagger_{-1-\sigma}f_{-1-\sigma})~~{\rm~for~all~}\sigma~~,
\ee 
where we used $c^2=id$.
This allows us to write:
\be
\label{J}
J=\prod_{\sigma\geq 0}{
\left[det(f_\sigma^\dagger f_\sigma)^{(-1)^{1-\sigma}}\right]}=
\prod_{\sigma=-N-1}^{N}{(detf^\dagger_\sigma f_\sigma)^{\frac{(-1)^{1-\sigma}}{2}}}=
\prod_{q=1-N}^{2+N}{
det(f^{(q)\dagger} f^{(q)})^{\frac{(-1)^{q}}{2}}}~~,
\ee
where we defined $f^{(q)}:=f_{1-q}:K^q\rightarrow H^{q}_d({\cal H})$.

Defining $b_\sigma:=dim_\C H_\sigma({\cal R})$ 
and $b^q:=dim_\C H^{q}({\cal H})$ (so that $b_\sigma:=b^{1-\sigma}$), let
us consider the `complex determinant line': 
\be
{\cal L}=det H^*_d({\cal H})=
\otimes_{\sigma=-N-1}^N{
\Lambda^{b_\sigma}[H_\sigma({\cal R})^{*(1-\sigma)}]}=\otimes_{q=1-N}^{2+N}
{\Lambda^{b^q}[H^q_d({\cal H})^{*q}]}~~,
\ee
which is a one-dimensional 
complex vector space. This space carries two Hermitian metrics. 
The first metric is induced by $g$ and $g_{\bf E}$ 
and arises upon transporting
the restriction of $h$ to the harmonic subspaces $K(\sigma)$ to metrics 
$h_\sigma=h^{(1-\sigma)}$ on $H_\sigma({\cal H})=H_d^{1-\sigma}({\cal H})$ 
through the Hodge isomorphisms 
$f_\sigma:K(\sigma)\rightarrow H_d^{1-\sigma}({\cal H})$: 
\be
\label{hsigma}
h_\sigma(u,v)=h(f_\sigma^{-1}u,f_\sigma^{-1}v)
=h^H_\sigma((f_\sigma^{-1})^\dagger f_\sigma^{-1}u,v)=
h^H_\sigma((f_\sigma f_\sigma^\dagger)^{-1}u, v) {\rm~~for~~} u,v\in 
H_d^{1-\sigma}({\cal H})~~.
\ee
The metrics (\ref{hsigma}) 
induce a metric $h_{\cal L}$ 
on ${\cal L}$ known as the {\em Hermitian} $L_2$-{\em metric}.
If $e^{(q)}_\alpha$ are bases for the complex vector spaces $H^q_d({\cal H})$, 
then $e:=\otimes_{q}{(e^{(q)}_1\wedge \dots \wedge e^{(q)}_{b^q})^{*q}}$ 
is a complex basis of ${\cal L}$, 
whose squared norm in the metric $h_{\cal L}$ is given by:
\be
||e||^2_{\cal L}=h_{\cal L}(e,e)=\prod_{q}{[detG^{(q)}]^{(-1)^q}}~~,
\ee
where $G^{(q)}_{\alpha\beta}=h^{(q)}(e^{(q)}_\alpha, e^{(q)}_\beta)$ are the (positive-definite) 
Hermitian Gramm matrices.
On the other hand, the auxiliary metrics $h_\sigma^H=h^{(1-\sigma)}_H$ on 
$H_\sigma({\cal H})=H{1-\sigma}_d({\cal H})$ 
induce a metric $h^H_{\cal L}$ on ${\cal L}$, for which:
\be
(||e||^H_{\cal L})^2=h^H_{\cal L}(e,e)=\prod_{q}{[detG_H^{(q)}]^{(-1)^q}}~~,
\ee
with $(G_H^{(q)})_{\alpha\beta}=h_H^{(q)}(e^{(q)}_\alpha, e^{(q)}_\beta)$.
It is clear from this description and from (\ref{J}), (\ref{hsigma}) that the two norms  
on ${\cal L}$ are related through:
\be
\label{norm_relation}
||.||^H_{\cal L}=J||.||_{\cal L}~~.
\ee
Let us consider the quantity:
\be
||.||_{RS}=T(L,A)||.||_{\cal L}~~,
\ee
which is a norm on ${\cal L}$ generalizing the standard Ray-Singer norm
\cite{BZ}. This norm is independent of $h^H_\sigma$.
With this definition, one can write ${\tilde Z}_{scl}$ as:
\be
{\tilde Z}_{scl}=\frac{||.||^H_{\cal L}}{~||.||_{RS}}~~,
\ee
which displays the dependence of ${\tilde Z}_{scl}$ on 
the auxiliary data $h^H_\sigma$.

\subsubsection{Metric induced on the real determinant line}

The traditional formulation of Ray-Singer torsion in the ungraded case involves connections 
defined on {\em real} vector bundles. As a consequence, 
one obtains a norm defined not on the complex determinant line
but on its real analogue. Since we work with complex vector bundles
and connections, it is more natural in our case 
to use the metric of the previous subsection, which 
is defined on the complex determinant line.  Here we explain 
the relation between this formulation and the traditional description.

Let us define the {\em real determinant line}: 
\be
\Lambda:=\otimes_{\sigma=-N-1}^N{
\Lambda^{2b_\sigma}_\R[H_\sigma({\cal R})^{*(1-\sigma)}]}=\otimes_{q=1-N}^{2+N}
{\Lambda_\R^{2b^q}[H^q_d({\cal H})^{*q}]}~~,
\ee
where all antisymmetrized and tensor products are now taken over the 
field of real numbers. 
There exists a natural isomorphism between $\Lambda$ and the 
second exterior power $\Lambda^2_\R {\cal L}$ of the complex line, 
where ${\cal L}$ is viewed as a real two-dimensional vector space 
by restriction of the field of scalars.
This isomorphism can be described as follows. Given bases 
$e^{(q)}_1\dots e^{(q)}_{b^q}$ of the complex vector spaces $H^q({\cal H})$, 
consider the bases $e^{(q)}_1\dots e^{(q)}_{b^q}, ie^{(q)}_1\dots ie^{(q)}_{b^q}$ of the 
underlying real vector spaces, as well as the associated bases 
$e:=\otimes_{q}{(e_1\wedge \dots\wedge e_{b^q})^{*q}}$ and 
$\epsilon:=\otimes_{q}{(e_1\wedge_\R \dots \wedge_\R e_{b^q}
\wedge_\R (ie_1)\wedge_\R \dots \wedge_\R (ie_{b^q}))^{*q}}$ of ${\cal L}$ and
$\Lambda$. Under a change $e^{(q)}_i\rightarrow e'^{(q)}_i=M_qe_i$ 
of the bases $e^{(q)}_1\dots e^{(q)}_{b^q}$ (where $M_q$ are invertible complex-
linear operators in $H^q({\cal H})$), we have:
\bea
e'&:=&\otimes_{q}{(e'_q\wedge \dots \wedge e'_q)^{*q}} =\prod_{q}{(det M_q)^{(-1)^q}}e~~\nn\\
\epsilon'&:=&\otimes_{q}{(e'_1\wedge_\R \dots \wedge_R e'_{b^q}
\wedge'_\R (ie'_1)\wedge_\R \dots \wedge_\R (ie'_{b^q}))^{*q}}=
|\prod_q{det M_q^{(-1)^q}}|^2\epsilon~~.
\eea
Considering the basis $e,ie$ of the real vector space ${\cal L}$, 
we obtain a basis $\epsilon_0:=e\wedge_\R (ie)$ of $\Lambda^2_\R{\cal L}$, 
which transforms as follows:
\be
\epsilon'_0:=e'\wedge_\R (ie')=|\prod_{q}{(det M_q)^{(-1)^q}}|^2 \epsilon_0~~.
\ee
The isomorphism $\phi:\Lambda^2_\R{\cal L}\rightarrow \Lambda$ 
is obtained by taking $\epsilon_0$ into $\epsilon$.
This is well-defined because $\epsilon$ and $\epsilon_0$ have the same 
transformation rules. Note that both of the real lines $\Lambda$ and $\Lambda^2_\R{\cal L}$ 
are equipped with orientations induced from the complex structure of 
$H^q({\cal H})$, and the vectors $\epsilon_0$ and $\epsilon$ always agree 
with these orientations.

Given the metrics $g$ on $L$ and $g_{\bf E}$ on 
${\bf E}$, we consider these constructions for 
bases $e_1\dots e_{b^q}$  which are orthonormal 
with respect to the Hermitian metrics $h^q=h_{1-q}$ which induced (through $f_\sigma$) on 
$H^q({\cal H})$.
In this case, only unitary transformations $M_q$ are allowed, hence 
$\epsilon_0$ and $\epsilon$ are uniquely determined by the metric data (since $|det M_q|=1$), 
while $e$ is determined up to multiplication by the phase factor $\prod_{q}{detM_q^{(-1)^q}}$. 
By construction, the Hermitian $L_2$ metric on ${\cal L}$ is uniquely determined by the 
constraint\footnote{Indeed, any element $v$ of ${\cal L}$ has the form 
$u=\alpha e$ with $\alpha$ a complex constant, so relation (\ref{norm0}) 
determines $||v||=|\alpha|$. Knowledge of the norm on ${\cal L}$ then 
determines the metric in standard fashion.}:
\be
\label{norm0}
||e||_{\cal L}=1~~.
\ee
This induces a Euclidean metric $(.,.)_{\cal L}:=Re h_{\cal L}( .,.)$ on 
the underlying real vector space, which makes $(e, ie)$ into a real 
orthonormal basis of ${\cal L}$. Upon taking the second exterior power, 
we have an induced Euclidean metric $(.,.)$ on $\Lambda_\R^2{\cal L}$, 
which is uniquely determined by the constraint:
\be
\label{norm1}
||\epsilon_0||=1~~.
\ee
In fact, given $u=\alpha e$ and $v=\beta e$ in ${\cal L}$, 
with $\alpha, \beta$ some complex constants, we have 
$u\wedge_\R v=Im({\overline \alpha}\beta) \epsilon_0$ and 
$||u\wedge_R v||=|Im({\overline \alpha}\beta)|$.

On the other hand, one has an {\em Euclidean $L_2$-metric} $(.,.)_\Lambda$
on $\Lambda$, which is uniquely determined by the condition:
\be
\label{norm2}
||\epsilon||_\Lambda=1~~.
\ee
This is the metric induced by the Euclidean scalar products
$(.,.)_\sigma:=Re h_\sigma(.,.)$ 
associated with the Hermitian metrics
$h_\sigma$ on $H_\sigma({\cal H})$. Since $\phi$ maps $\epsilon_0$ into 
$\epsilon$, it is clear from (\ref{norm1}) and (\ref{norm2}) that 
$||\phi(v)||=||v||_\Lambda$. Hence the Euclidean $L_2$ metric is naturally 
induced by the Hermitian $L_2$ metric via the (metric-independent) 
isomorphism $\phi$. 

Conversely, the Euclidean $L_2$ metric determines the 
vector $\epsilon$ (and thus the vector $\epsilon_0$), via relation 
(\ref{norm2}). (The apparent sign ambiguity is fixed by 
the condition that $\epsilon$ must agree with the orientation of $\Lambda$).
Then $e$ is determined up to a phase factor by the relation 
$e\wedge_\R (ie)=\epsilon_0$. Since multiplying $e$ by a 
phase factor does not 
affect the metric determined by relation (\ref{norm0}), it follows 
that $||.||_{\cal L}$ is uniquely determined by the 
Euclidean $L_2$ metric.

We conclude that the Hermitian and Euclidean $L_2$ metrics completely 
determine each other. Which one we use is simply a matter of convention. 
In this note, we use the metric $||.||_{\cal L}$, which is better adapted to 
our complex formalism. In the ungraded case $({\bf E}=E_0$), 
the Euclidean $L_2$ metric arises when considering the complex bundle
as a real bundle by forgetting its complex structure.

\subsubsection{Metric-independence of the Ray-Singer norm}

It turns out that the generalized Ray-Singer norm is {\em independent} 
of $g$ and $g_{\bf E}$, and therefore completely independent of metric
data. This statement, which parallels a well-known result for the ungraded 
case due to Ray and Singer,  follows from the general discussion given in 
\cite{Schwarz_resolvent} and \cite{Adams_Sen}. 
The results of \cite{Adams_Sen} assure independence of $||.||_{RS}$
of all metric data due to the following facts:

\

(1) The resolvent (\ref{res}) is {\em topological}, i.e. the complex 
\be
~~0\rightarrow{\cal H}(N)
\stackrel{T_N=d}{\longrightarrow}\dots 
\stackrel{T_2=d}{\longrightarrow} 
{\cal H}(1)\stackrel{T_1=d}{\longrightarrow}ker(d_{s=0})=ker(T_0)~~
\ee
is defined without reference to any metric.

\

(2) The resolvent is also {\em elliptic}, i.e. the deformed Laplacians
$\Delta_\sigma=T^t_\sigma T_\sigma+T_{\sigma+1}T_{\sigma +1}^t=
(dd^\dagger +d^\dagger d)_{s=\sigma}$
are elliptic operators for all $\sigma=0\dots N$.

\

(3) The base manifold $L$ is odd dimensional. 

\

The proof (which can be found in \cite{Adams_Sen}) 
involves a combination of results in elliptic 
operator theory with arguments in linear algebra.

A particularly simple case arises when the background superconnection $A$
is {\em acyclic}, i.e. the cohomology $H_d^*({\cal H})$ 
vanishes in all degrees. In this situation, the quantity 
${\tilde Z}_{scl}$ coincides with $T(L, A)^{-1}$ and
 is independent of all metric data. Examples of acyclic backgrounds are 
provided by condensation of scalars in boundary condition changing sectors 
of `topological brane-antibrane pairs', as discussed in detail in 
\cite{bv} and \cite{gauge}. 

\subsubsection{The complex prefactor}

Let us turn to the complex prefactor $C$ given in (\ref{C}).
A result of \cite{Adams_Sen} implies that:
\be
\zeta_{|T_0|}(0)=-\sum_{\sigma=0}^N{(-1)^\sigma dim H_\sigma({\cal R})}=
\sum_{q=1-N}^1{(-1)^q dim H^q_d({\cal H})}~~,
\ee
which shows that the absolute value of $C$ is independent of metric data. \
This shows that $|Z_{scl}|$ depends only on the metrics $h^H_\sigma$.

On the other hand, the value $\eta_{T_0}(0)$ may depend on $g$ and 
$g_{\bf E}$, 
which means that the phase factor of the semiclassical partition function 
will generally carry a metric dependence. This parallels well-known results 
valid for the ungraded case (standard Chern-Simons field theory)
\cite{Witten_tcs}.

\section{Example: D-brane pairs of unit relative grade 
in a scalar background}

Consider a D-brane pair $(a,b)$ such that $grade(a)=0$ and $grade(b)=1$. 
In this case, the underlying graded bundle is ${\bf E}=E_a\oplus E_b$, where 
$E_a$ and $E_b$ are the flat bundles describing the reference 
D-brane background. We let $A_a$ and $A_b$ be associated flat connections.
The space ${\cal H}^q$ consists of sections of the bundle
${\cal V}^q=\Lambda^q(T^*L)\otimes End(E_a)\oplus \Lambda^q(T^*L)
\otimes End(E_b)\oplus \Lambda^{q-1}(T^*L)
\otimes Hom(E_a, E_b)\oplus \Lambda^{q+1}(T^*L)\otimes Hom(E_b, E_a)$. 
Such elements can be viewed as matrices of bundle-valued forms:
\be
\label{u}
u=\left[\begin{array}{cc}
u_q&u_{q+1}\\
u_{q-1}&{\hat u}_q
\end{array}\right]~~,~~{\rm for}~~|u|=q~~,
\ee
where the subscript denotes form rank and the bundle 
components of $u_q=u_{aa}, {\hat u}_q=u_{bb}, u_{q+1}=u_{ba}$ and 
$u_{q-1}=u_{ab}$ satisfy $u_{\alpha\beta}\in 
\Omega^{q+grade(\alpha)-grade(\beta)}(L, Hom(E_\alpha, E_\beta))$. 
This corresponds to presenting $u$ as the direct sum:
\be
u=\oplus_{\alpha,\beta\in \{a,b\}}{u_{\alpha\beta}}~~.
\ee

We shall consider backgrounds of the form:
\be
\label{phi}
\phi=\left[\begin{array}{cc} 0&0\\\phi_0&0\end{array}\right]~~,~~|\phi|=1~~,
\ee
where $\phi_0$ is a
zero-form valued in the bundle $Hom(E_a, E_b)$. In this case, the 
equations of motion $d\phi+\phi\bullet \phi=0$ reduce to $d\phi_0=0$, which 
means that $\phi_0$ is a covariantly-constant section of $Hom(E_a,E_b)$. 
The shifted background is the flat superconnection 
$A=(A_a\oplus A_b)+\phi$. We shall assume that $\phi_0$ is a 
bundle morphism in the restricted sense, i.e. we require\footnote{This assumption 
allows us to treat the kernel and cokernel of $\phi_0$ as subbundles of 
$E_a$ and $E_b$. Allowing maps $\phi_0$ of non-constant rank leads to situations
which are better described in terms of sheaf theory. This is very similar to the case of holomorphic 
bundles vs. holomorphic sheaves. Note that constancy of $rk \phi_0(p)$ 
is a purely technical assumption -- there is no {\em physical} 
reason to restrict to scalar backgrounds of constant rank.} that
the rank of the fiber maps $\phi_0(p):(E_a)_p\rightarrow (E_b)_p$ 
is independent of the point $p\in L$. 

Let us assume that the reference flat bundles $(E_a, A_a)$ and $(E_b, A_b)$ 
are unitary, i.e. they admit covariantly-constant Hermitian metrics
\footnote{In 
physical language, this means that we are dealing with unitary connections, 
i.e. connections whose matrices are anti-Hermitian in appropriate 
local frames.}.
We shall pick the metric $g_{\bf E}$
on ${\bf E}$ to be induced by two such metrics. 
With this hypothesis, 
it was showed in \cite{gauge} that the deformed Laplacian 
$\Delta_\phi=d_\phi d_\phi^\dagger+
d_\phi^\dagger d_\phi$ in the  background $\phi_0$ has the form:
\be
\label{DD}
\Delta_\phi u=
\left[\begin{array}{cc}
\Delta_\phi^{(aa)}u_q&\Delta_\phi^{(ba)}u_{q+1}\\
\Delta_\phi^{(ab)}u_{q-1}&\Delta_\phi^{(bb)}{\hat u}_q
\end{array}\right]~~,
\ee
with:
\bea
\Delta_\phi^{(\alpha\beta)}u_{\alpha\beta}&=&\Delta u_{\alpha\beta} +
u_{\alpha\beta}\circ t^{(\alpha)}+
t^{(\beta)}\circ u_{\alpha\beta}~~,
\eea
where we defined:
\be
t^{(a)}=\phi_0^\dagger \phi_0\in End(E_a)~~{\rm~and~}~~t^{(b)}=\phi_0\phi_0^\dagger \in End(E_b)~~
\ee
and where $\Delta u_{\alpha\beta}$ stands for the form 
Laplacian coupled to the flat connection induced by $A_\alpha$ and $A_\beta$ 
on the bundle $Hom(E_\alpha,E_\beta)$. In direct sum notation, we have:
\be
\Delta_\phi=\oplus_{\alpha\beta}{\Delta_\phi^{(\alpha\beta)}}~~.
\ee
The fact that the deformed Laplacian decomposes in this manner is a characteristic of 
scalar background of the particular form (\ref{phi}).

Since the subspaces $\Omega^*(L, Hom(E_\alpha,E_\beta))\subset {\cal H}$  
are mutually orthogonal with respect to the scalar product $h$, we have:
\be
p_q:=det^{',reg}_{|.|=q}\Delta_\phi=
det^{',reg}_{rk =q}\Delta_\phi^{(aa)}
det^{',reg}_{rk =q}\Delta_\phi^{(bb)}
det^{',reg}_{rk =q-1}\Delta_\phi^{(ab)}
det^{',reg}_{rk =q+1}\Delta_\phi^{(ba)}~~
\ee
and:
\be
T(L,A)^2=\prod_{q=1-N}^{N+2}{(det^{',reg}_{|.|=q}\Delta_\phi)^{q(-1)^q}}=
\prod_{q=1-N}^{N+2}{p_q^{q (-1)^q}}~~.
\ee

It was showed in \cite{gauge} that the kernel $K$ of $\Delta_\phi$ 
coincides (up to a twist of gradings) 
with the space of harmonic forms valued in the 
flat bundle $End(R\oplus I^\perp)$, where $R=ker \phi$ and 
$I=im \phi$ are (flat) subbundles of $E_a$ and $E_b$:
{\scriptsize \be
K^q=\Omega^q_{harm}(L, End(R))\oplus \Omega^q_{harm}(L, End(I^\perp))\oplus 
\Omega^{q-1}_{harm}(L, Hom (R, I^\perp))\oplus \Omega^{q+1}_{harm}
(L, Hom(I^\perp, K))~~.
\ee}
Using the covariantly-constant metrics 
on $E_a$ and $E_b$, we identify $I^\perp$ with 
the cokernel $Q$ of the flat bundle map $\phi_0:E_a\rightarrow E_b$:
\be
I^\perp\approx Q:=coker \phi_0=E_b/im \phi_0~~.
\ee
Hodge theory leads to identifications $H^q_{d_\phi}({\cal H})=K^q$ and 
$\Omega^k_{harm}(V)\approx H^k(V)$, where $V$ is any of the 
flat bundles $End(R)$, $End(I^\perp)$, $Hom(R, I^\perp)$ and 
$Hom(I^\perp, R)$, while $H^*(V)$ stands for the usual cohomology with 
coefficients in the local system determined by $V$. Combining everything, 
we obtain:
{\scriptsize \bea
\label{Hd}
H^q_{d_\phi}({\cal H})=H^q(L, End(R))\oplus H^q(L, End(Q))\oplus 
H^{q-1}(L, Hom (R, Q))\oplus H^{q+1}(L, Hom(Q, R))~~.
\eea}
Using (\ref{Hd}), we find that the 
Ray-Singer metric is defined on the determinant line:
{\footnotesize \bea
{\cal L}&=&det H^*_{d_\phi}({\cal H})=
\otimes_q{\Lambda^{max}H^q_{d_\phi}({\cal H})^{* q}}=\\
&=&det H^*(L, End(R))\otimes det H^*(L,End(Q))
\otimes [det H^*(L,Hom(R, Q))]^*\otimes [det H^*(L, Hom(Q, R))]^*~~.\nn
\eea}

\subsection{The case of trivial flat bundles}

Let us consider the particularly simple 
case when $E_a$ and $E_b$ (of ranks $r_a$ and $r_b$) 
are trivial as flat vector 
bundles, so that all components $u_k, {\hat u}_k, u_{k+1}, u_{k-1}$ can be 
viewed as operator-valued forms, and $d$ coincides with the de Rham 
differential acting on such forms. In this situation, the condition 
$d\phi_0=0$ means that $\phi_0$ is a {\em constant} linear operator from 
$\C^{r_a}$ to $\C^{r_b}$. Since the flat bundle structure is trivial, we 
expect to obtain a particularly simple expression for $T(L,A)$. 
We show below that this expectation is fulfilled. 

For this, we first notice that the non-negative operators
$t^{(a)}=\phi_0^\dagger \phi_0\in End(\C^{r_a})$ 
and $t^{(b)}=\phi_0\phi_0^\dagger\in End(\C^{r_b})$ have the same nonzero eigenvalues. 
This follows form the following variant of the argument used in Subsection 
3.4.2. Noting that $ker (\phi_0^\dagger \phi_0)^\perp=ker (\phi_0)^\perp$ 
and $ker (\phi_0\phi_0^\dagger)^\perp=ker (\phi_0^\dagger)^\perp
=im(\phi_0) $, the restriction of $\phi_0$ gives an isomorphism:
\be
\phi_0:ker(\phi_0^\dagger \phi_0)^\perp\stackrel{\approx}{\rightarrow}
ker (\phi_0\phi_0^\dagger)^\perp~~.
\ee
This induces a bijection between the non vanishing eigenvalues of 
$\phi_0^\dagger\phi_0$ and $\phi_0\phi_0^\dagger$, since 
$\phi_0^\dagger \phi_0 v=\lambda v$ for some $\lambda\in \R_+^*$ 
and $v\in  \C^{r_a}$
implies $\phi_0\phi_0^\dagger(\phi v)=\lambda \phi_0 v$. 
In particular, the ranks of the two operators coincide:
\be
rk(\phi_0^\dagger \phi_0)=rk(\phi_0 \phi_0^\dagger):=\rho~~,
\ee
while their defects are given by:
\be
dim~ker (\phi_0^\dagger \phi_0)=rk R=r_a-\rho~~,~~
dim~ker (\phi_0 \phi_0^\dagger)=rk Q=r_b-\rho~~.
\ee

Consider unitary transformations $S_a$ and $S_b$ of $\C^{r_a}$ and $\C^{r_b}$ 
which diagonalize $\phi_0^\dagger \phi_0$ and $\phi_0\phi_0^\dagger$:
\bea
\label{S_tf}
&&S_a\phi_0^\dagger \phi_0 S_a^{-1}=D^{(a)}:=diag(\lambda_1^{(a)}\dots 
\lambda^{(a)}_{r_a})~~\nn\\
&&S_b\phi_0 \phi_0^\dagger S_b^{-1}=D^{(b)}:=
diag(\lambda^{(b)}_1\dots \lambda^{(b)}_{r_b})~~,
\eea
with $\lambda_i^{(a)}, \lambda^{(b)}_j\geq 0$.
In view of the above, we can always pick $S_a$ and $S_b$ such that:
\bea
\lambda^{(a)}_i&=&\lambda^{(b)}_i:=\lambda_i~{\rm~for~}~i=1\dots 
\rho~~{\rm and}\nn\\
\lambda^{(a)}_i&=&0~{\rm~for~}~i=\rho+1 \dots r_a~~\\
\lambda^{(b)}_j&=&0~{\rm~for~}~j=\rho+1 \dots r_b~~.\nn
\eea

It is easy to see that the unitary transformation 
$S=S_a\oplus S_b\equiv \left[\begin{array}{cc}S_a&0\\0&S_b\end{array}\right]$
preserves the decomposition (\ref{DD}) of $\Delta_\phi$ 
while bringing $\Delta_\phi^{(\alpha\beta)}$ to the form:
\be
\Delta_\phi^{(\alpha\beta)}u_{\alpha\beta} = 
\Delta u_{\alpha\beta}+D^{(\beta)}\circ u_{\alpha\beta} +
u_{\alpha\beta} \circ D^{(\alpha)}~~,
\ee
which act on the matrix components\footnote{Picking orthonormal 
bases $e_i^{(a)}$ and $e_j^{(b)}$ of $\C^{r_a}$ and $\C^{r_b}$, we
define $u_{\alpha\beta}^{ji}$ through $u_{\alpha\beta}(e_i^{(\alpha)})=
u_{\alpha\beta}^{ji}e_j^{(\beta)}$.} of $u_{\alpha\beta}$ as:
\be
\Delta_\phi^{(\alpha\beta)}u^{ji}_{\alpha\beta} = \Delta u^{ji}_{\alpha\beta}+
(\lambda^{(\alpha)}_i+\lambda^{(\beta)}_j)u^{ji}_{\alpha\beta}~~.
\ee
This leads to the expression:
\be
p_q=p_q^{(1)}p_q^{(2)}p_q^{(3)}~~,
\ee
where:
\bea
p_q^{(1)}&=&det^{',reg}_{rk=q}(\Delta)^{(r_a-\rho)^2+(r_b-\rho)^2}
\left[det^{',reg}_{rk=q-1}(\Delta)
det^{',reg}_{rk=q+1}(\Delta)\right]^{(r_a-\rho)(r_b-\rho)}~~\\
p_q^{(2)}&=&
\prod_{i,j=1}^\rho{\left[
det^{',reg}_{rk=q}(\Delta +\lambda_i+\lambda_j)\right]^2
det^{',reg}_{rk=q-1}(\Delta +\lambda_i+\lambda_j)
det^{',reg}_{rk=q+1}(\Delta +\lambda_i+\lambda_j)}~~\nn\\
p_q^{(3)}&=&\prod_{i=1}^{\rho}{
det^{',reg}_{rk=q}(\Delta +\lambda_i)^{2(r_a+r_b-2\rho)}
\left[det^{',reg}_{rk=q-1}(\Delta +\lambda_i)
det^{',reg}_{rk=q+1}(\Delta +\lambda_i)\right]^{r_a+r_b-2\rho}}~~.\nn
\eea
It is easy to check that: 
\be
\prod_q{\left[det^{',reg}_{rk=q-1}(\Delta +\mu)
det^{',reg}_{rk=q+1}(\Delta +\mu)\right]^{q(-1)^q}}=
\left(\prod_q{\left[det^{',reg}_{rk=q}(\Delta +\mu)\right]^{q(-1)^q}}\right)
^{-2}~~.
\ee
Therefore, one has
$\prod_{q}{(p_q^{(2)})^{q(-1)^q}}=
\prod_{q}{(p_q^{(3)})^{q(-1)^q}}=1$ and we obtain:
\be
\label{final}
T(L,A)=\left[\prod_q{(p_q)^{q(-1)^q}}\right]^{1/2}=
\left[\prod_q{(p_q^{(1)})^{q(-1)^q}}\right]^{1/2}=
T(L)^{(r_a-r_b)^2}
~~.
\ee
where $T(L)$ is the usual Ray-Singer torsion of $L$
(i.e. the standard analytic torsion for the trivial flat complex line bundle 
${\cal O}_L$ over $L$):
\be
T(L)=\prod_q{det^{',reg}_{rk=q}(\Delta)^{\frac{q(-1)^q}{2}}}=
e^{-\frac{1}{2}\sum_{q=0}^3{(-1)^{q} q \zeta'_{\Delta_q'}(0)}}~~,
\ee
with $\Delta_q$ the Laplacian on $\Omega^q(L)$.  
The determinant line can be found by noticing that 
$R={\cal O}_L^{\oplus (r_a-\rho)}$ and $Q={\cal O}_L^{\oplus (r_b-\rho)}$,
where ${\cal O}_L$ is the trivial complex flat line bundle over $L$. 
This implies 
$End(R)={\cal O}_L^{\oplus (r_a-\rho)^2}$, 
$End(Q)={\cal O}_L^{\oplus (r_b-\rho)^2}$
and 
$Hom(R,Q)=Hom(Q,R)={\cal O}_L^{\oplus (r_a-\rho)(r_b-\rho)}$
(as flat line bundles), thereby giving the determinant line:
\be
{\cal L}=det H^*(L, {\cal O}_L)^{\otimes (r_a-r_b)^2}=
det H^*(L)^{\otimes (r_a-r_b)^2}~~.
\ee
Combining with (\ref{final}), we see that $T(L,A)$ is the norm induced 
on $det H^*(L, {\cal O}_L)^{\otimes (r_a-r_b)^2}$ by the 
usual Ray-Singer norm on $det H^*(L, {\cal O}_L)=det H^*(L)$ (considered
in the `complex' formalism discussed in Subsection 3.4.3).

{\bf Observation} 
If $r_a=r_b$, then it was showed in 
\cite{bv, gauge} that a background $\phi_0$ which is a flat isomorphism 
leads to an acyclic composite, i.e. the complex 
of the shifted differential $d_\phi$ is acyclic in all degrees. 
In this case (and with trivial flat connections $A_a$ and $A_b$), 
the generalized Ray-Singer torsion is equal to one. 

\acknowledgments{I thank R. Roiban for comments on the manuscript and 
Martin Rocek for interest and support.
The present work was supported by the Research Foundation under NSF 
grant PHY-9722101.}

\

\end{document}